\documentclass[sn-mathphys,Numbered]{sn-jnl}

\usepackage{graphicx}%
\usepackage{multirow}%
\usepackage{amsmath,amssymb,amsfonts}%
\usepackage{amsthm}%
\usepackage{mathrsfs}%
\usepackage[title]{appendix}%
\usepackage{xcolor}%
\usepackage{textcomp}%
\usepackage{manyfoot}%
\usepackage{booktabs}%
\usepackage{algorithm}%
\usepackage{algorithmicx}%
\usepackage{algpseudocode}%
\usepackage{listings}%
\usepackage{lineno}

\theoremstyle{thmstyleone}%
\theoremstyle{thmstyletwo}%

\theoremstyle{thmstylethree}%

\begin{document}

\title[Probability of an AMOC Collapse]{Probability Estimates of a 21$^{\mathrm{st}}$ Century AMOC Collapse} 


\author*[1]{\fnm{Emma J.V.} \sur{Smolders}}\email{e.j.v.smolders@uu.nl}

\author[1]{\fnm{Ren\'e M.} \spfx{van} \sur{Westen}}\email{r.m.vanwesten@uu.nl}

\author[1]{\fnm{Henk A.} \sur{Dijkstra}}\email{h.a.dijkstra@uu.nl}

\affil[1]{\orgdiv{Department of Physics}, \orgname{Institute for Marine and Atmospheric research Utrecht, Utrecht University}, \orgaddress{\street{Princetonplein 5}, \city{Utrecht}, \postcode{3584 CC},  \country{the Netherlands}}}


\abstract{{\bf There is increasing concern that the Atlantic Meridional Overturning Circulation (AMOC) may collapse this century  with a disrupting societal impact on large parts of the world. Preliminary estimates of the probability of such an AMOC collapse have so far been  based on conceptual models and statistical analyses of proxy data. Here, we provide  observationally based estimates of such probabilities from reanalysis data. We first identify optimal observation regions of an AMOC collapse from a recent global climate model simulation. Salinity data near the southern boundary of the Atlantic turn out to be optimal to provide estimates of the time of the AMOC collapse in this model. Based on the reanalysis products, we next determine probability density functions of the AMOC collapse time. The collapse time is estimated between 2037-2064 (10-90\% CI) with a mean of 2050 and the probability of an AMOC collapse before the year 2050 is estimated to be $59 \pm 17\%$.} 
}




\maketitle

The Atlantic Meridional Overturning Circulation (AMOC) transports relatively warm surface waters northward and cold deep 
waters southward, thereby maintaining Western Europe's mild climate and strongly modulating global climate patterns 
\cite{Rahmstorf2002}.  The AMOC is becoming an ever more studied component of the climate system as it is considered 
one of the major tipping systems which may undergo a transition under anthropogenic climate change \cite{Armstrong2022}. 
The AMOC can potentially collapse as a consequence of surface freshwater input in the North Atlantic, e.g. ice melt from the 
Greenland Ice Sheet or a change in surface freshwater fluxes.  A collapse from its current strong northward overturning state 
to a substantially weaker or reversed state would have major climate impacts such as  a meridional shift 
in the tropical rain belts, dynamical sea-level changes, and a substantial cooling in Northwestern Europe 
\cite{Orihuela2022, vanWesten2024}.  Evidence of past AMOC changes comes from paleoclimatic reconstructions, which 
suggest an alternation  between stronger and weaker states during the Dansgaard-Oeschger events \cite{Ganopolski2001, 
Vettoretti2022}. Determining the probability of such a transition to happen before the year~2100 is therefore an urgent problem 
in climate research. 

The AMOC has been monitored along the RAPID transect at 26$^\circ$N since 2004 \cite{Frajka-Williams2019AtlanticVariability}, 
along the SAMBA transect at 34.5$^\circ$S since 2009 and along the OSNAP transect 
spanning from 53$^\circ$N to 60$^\circ$N since 2014. Because the direct observational record of RAPID is only 20~years long, historical  
AMOC reconstructions have been developed using sea surface temperature (SST) observations over the sub-polar gyre. These so-called AMOC fingerprints  \cite{Ceasar2018} indicate a weakening of the AMOC by 3 $\pm$ 1 Sv since 1950.  
Using these fingerprints, recent studies \cite{Boers2021,  Ditlevsen2023} have used statistical indicators, referred to 
as Early Warning Signals (EWS),  to investigate the proximity of the  AMOC to its collapse. 

The classical EWS are based on critical slowdown, which is expected near a saddle-node bifurcation \cite{Ditlevsen2010}, 
and consist of  a lag-1 autocorrelation tending to unity and an increase in variance.  Ditlevsen \& Ditlevsen (2023) 
\cite{Ditlevsen2023} estimated that the present-day AMOC would collapse in the year 2057 with 2025 and 2095 as 
the 95\% confidence values.   Because of the many assumptions behind this estimate, both regarding the 
proxy-data based AMOC reconstruction used and the statistical methodology, the 
results have received substantial  criticism \cite{Ben-Yami2023}.  One can also determine the 
probability of an AMOC collapse in models by using rare-event algorithms \cite{Rolland2018}.  However, because 
of the  computational complexity such methodology has so far only been applied to rather idealised AMOC models 
\cite{Castellana2019, Castellana2020, Baars2021, Cini2024, vanWesten2024c}.  

Our novel approach here to determine probability estimates of an AMOC collapse before the year~2100 
starts from recent modelling work  \cite{vanWesten2024} that  has shown that an AMOC collapse does occur 
in the CMIP5 version of the Community Earth System Model (CESM).  It was shown for this model that the 
classical EWS using the sub-polar gyre SSTs do not give an alarm for  the AMOC collapse  \cite{vanWesten2024}.  
This result motivates  to determine the optimal regions and observables that can predict the AMOC 
tipping time in the  CESM. We identify these regions using additional CESM simulations and then
use reanalysis data to estimate the distribution of the present-day AMOC tipping time from 
observations.   

\section*{Optimal Observation Locations}

\begin{figure}[h!]

\includegraphics[width=1\columnwidth, trim = {0cm 0cm 0cm 0cm}, clip]{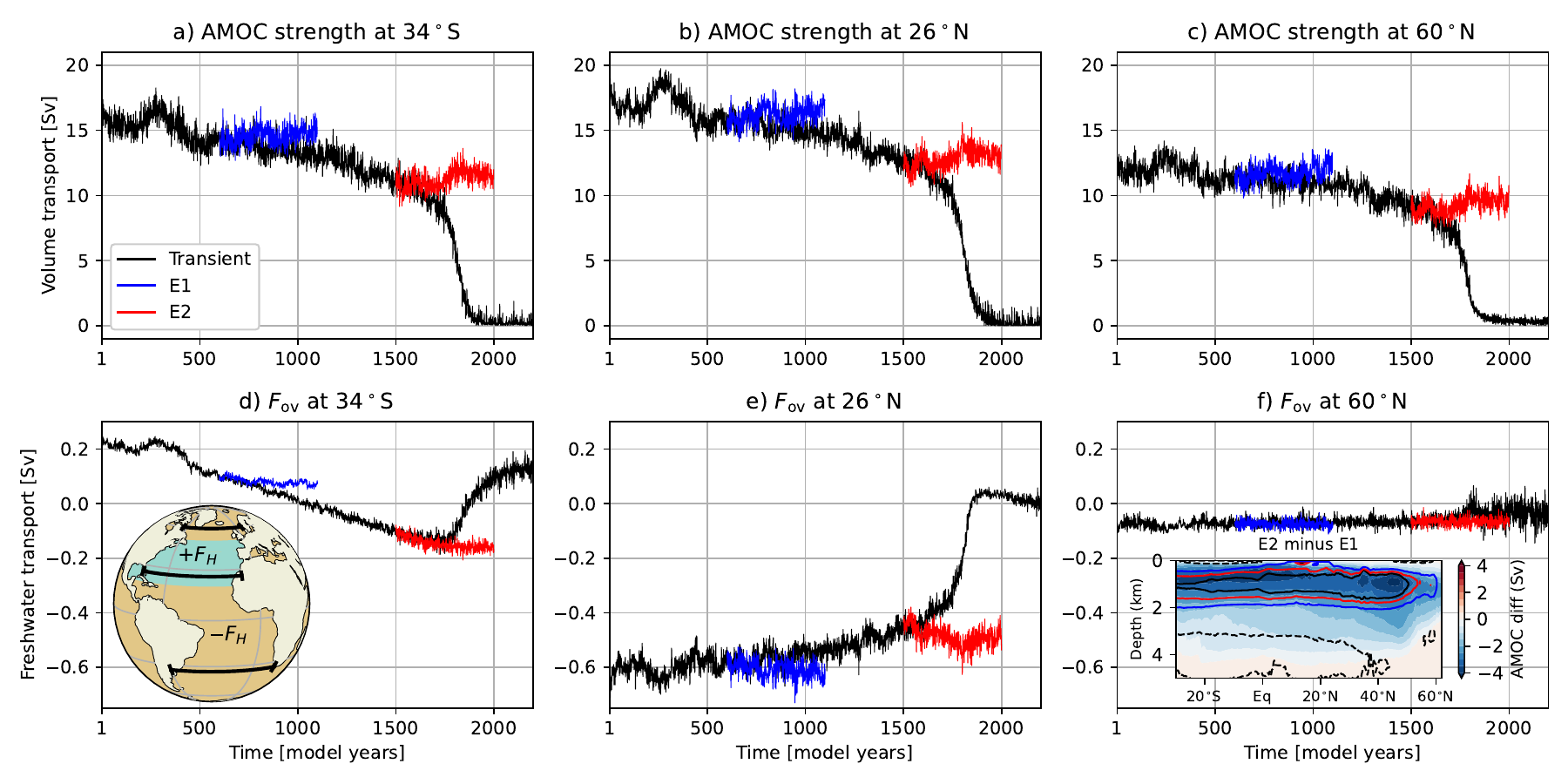}

\caption{\textbf{Volume and AMOC induced freshwater transports at the observational transects in the CESM.}
(a -- c): Maximum AMOC strength and (d -- f): freshwater transport by the overturning component $F_{\mathrm{ov}}$ 
of the quasi-equilibrium (black) and  equilibrium CESM simulations (blue for $E_1$ and red for $E_2$) at 34$^\circ$S, 
26$^\circ$N and 60$^\circ$N.  Note that the maximum of the AMOC strength is found below 500~m. The inset in panel~d 
shows the location  where surface fresh water is added between 20$^\circ$N and 50$^\circ$N in the Atlantic Ocean (cyan); 
this is globally compensated (brown).  The black horizontal lines indicate the three transects.  The inset in panel~f shows 
in color the difference in the AMOC ($E_2$ minus $E_1$, last 50~years);  the contours indicate the AMOC stream function  
of $E_1$  (black = 14 Sv, red = 12 Sv, blue = 9 Sv, black dashed = -1 Sv).} 

\label{fig:Figure_1}
\end{figure}

Our starting point is the pre-industrial quasi-equilibrium CESM simulation  \cite{vanWesten2024}   where a surface 
freshwater anomaly $F_H$  is added in the North Atlantic (inset in Figure~\ref{fig:Figure_1}d) with a rate of 
$3 \times 10^{-4}$ Sv/year  (see Methods).  The maximum AMOC strength at different latitudinal sections near 
to the observational array transects in the Atlantic ($34^\circ$S,  $26^\circ$N, and $60^\circ$N) for this simulation 
are shown in Figures~\ref{fig:Figure_1}a,b,c  (black curves),  respectively. The gradual increase in surface freshwater 
forcing  results in a rapid 
decrease in AMOC strength around model year~1758, with a difference of about 8~Sv at 
$26^\circ$N over 100~model years. The  AMOC-induced freshwater transport (indicated  as $F_{\mathrm{ov}}$, see Methods) at 
34$^\circ$S has been proposed as a physics-based early warning indicator for AMOC stability \cite{vanWesten2024}. 
This indicator goes through a  minimum just before the collapse (Figure~\ref{fig:Figure_1}d). The $F_{\mathrm{ov}}$ 
value at 26$^\circ$N  (Figure~\ref{fig:Figure_1}e) increases strongly  through the transition but  the 
$F_{\mathrm{ov}}$ value at  60$^\circ$N (Figure \ref{fig:Figure_1}f) remains fairly constant.

To determine optimal regions for the prediction of an AMOC collapse in this CESM simulation,  we branched off two simulations from the 
quasi-equilibrium simulation \cite{vanWesten2024c}. The two simulations have a different but constant freshwater forcing 
$F_H$ and we refer  to them as $E_1$ ($F_H = 0.18$~Sv, blue curves branched off from model year~600) and $E_2$ ($F_H = 0.45$~Sv, 
red curves branched off from model year~1500);  both simulations are integrated for 500~years.  The simulations $E_1$ and $E_2$ equilibrate 
after about 300~years and are almost in statistical equilibrium in the remaining part of the simulation.  The value of  $F_{\mathrm{ov}}$ at 34$^{\circ}$S for $E_1$ is 0.06~Sv (last 50~years) which 
is close to the quasi-equilibrium simulation value of 0.10~Sv (model years~575 -- 625). Values of $F_{\mathrm{ov}}$ for $E_2$ 
drift away a bit more, with a time mean at 34$^\circ$S of -0.16~Sv (last 50~years), compared to 
-0.10~Sv in the quasi-equilibrium simulation (model years~1475 -- 1525). The $F_{\mathrm{ov}}$ values at 26$^{\circ}$N and 
60$^{\circ}$N for $E_1$ and $E_2$ remain very close to their branching point values.  As $F_{\mathrm{ov}}$ at 34$^{\circ}$S is considered an
important indicator for AMOC stability \cite{vanWesten2024}, the simulation $E_2$ has a lower value of $F_{\mathrm{ov}}$ at 34$^{\circ}$S and 
is closer to the tipping point than  $E_1$.  Hence we expect  that (classical) EWS  would show stronger signals of critical 
slowdown in $E_2$ than in $E_1$. 

To quantify the ratio of the EWS calculated from  $E_2$ and $E_1$, we use yearly averaged and linearly detrended 
data from model years~350 -- 500 where the simulations are best equilibrated.  A sliding window of 70 years is used 
(the results are robust when using a  window of 60-80 years) and the average ratio over all possible permutations is 
computed according to: 
\begin{equation}
    R^X_I = \frac{1}{(N-M)^2} \sum_{i=0}^{N-M}\sum_{j=0}^{N-M}\frac{I(X_{E_2}[i:i+M])}{I(X_{E_1}[j:j+M])}
    \label{eq:mean_sigma}
\end{equation}
where $N = 150$~years, $I$ indicates the type of  EWS, $M$ 
the sliding  window size (70~years), and $X$  either the temperature 
($T$) or the salinity ($S$). 
Locations with $R^X_I > 1$ for  $I = \mathrm{VAR}$ (variance) and 
$I = \mathrm{AC1}$ (lag-1 autocorrelation) are signatures of a stronger critical slowdown in $E_2$.  
The null-hypothesis that $R^X_I = 1$ is rejected when the probability $p(R^X_I > 1) > 0.9$. Furthermore, 
a requirement that the value of the $\mathrm{AC1} > 0.5$ is applied in order to avoid large values of 
$R^X_{\mathrm{AC1}}$ due to low $\mathrm{AC1}$ values.   

\begin{figure}[h!]

\includegraphics[width=1\columnwidth, trim = {0cm 0cm 0cm 0cm}, clip]{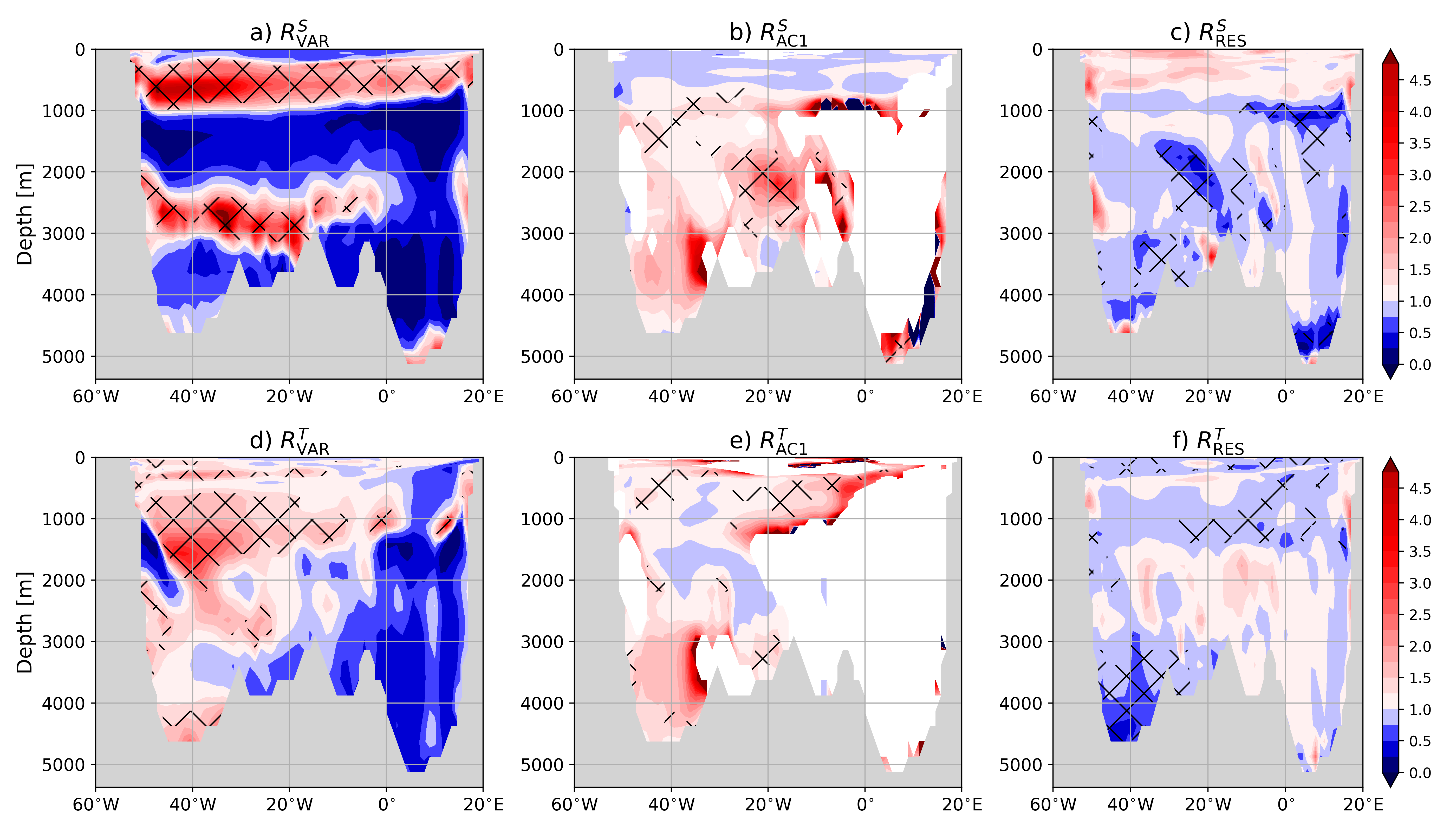}

\caption{\textbf{Early Warning Signals along 34$^{\circ}$S.}
(a -- c): Average  ratio of salinity-based EWS indicators $R^S_{\mathrm{VAR}}, R^S_{\mathrm{AC1}}$ and $R^S_{\mathrm{RES}}$ and
(d -- f): temperature-based indicators $R^T_{\mathrm{VAR}}, R^T_{\mathrm{AC1}}$ and $R^T_{\mathrm{RES}}$ at the SAMBA 
transect (34$^{\circ}$S). A sliding  window of 70 years is used and a significance level of 0.9 is  indicated by the hatched areas.
Missing values in the $\mathrm{AC1}$ plots are due to the restriction that $\mathrm{AC1} > 0.5$.}

\label{fig:Figure_2}
\end{figure}

Values of $R^X_{\mathrm{VAR}}$ and  $R^X_{\mathrm{AC1}}$ are shown in Figure \ref{fig:Figure_2} 
along the SAMBA transect at 34$^{\circ}$S for salinity (panels a,b) and temperature (panels d,e). 
The quantity  $R^S_{\mathrm{VAR}}$ distinctly shows two bands of significant values, one 
just below the surface ranging to 1000~m (i.e., Atlantic Surface Water and Antarctic Intermediate Water), 
and the other on the western part of the transect ranging between 2000~m and 3000~m depth (i.e., North 
Atlantic Deep Water).  The value of $R^T_{\mathrm{VAR}}$ indicates a significantly larger variance in 
$E_2$ compared to  $E_1$ in almost the entire western part of the transect.  The values of $R^S_{\mathrm{AC1}}$ 
and $R^T_{\mathrm{AC1}}$ predominantly show significance in the western part,  while the eastern part 
is characterised by too low AC1 values.  

One would expect that classical EWS (i.e., variance and lag-1 autocorrelation) consistently identify regions 
where the AMOC in CESM shows signs of critical slowdown. There is indeed some overlap 
between significant $\mathrm{VAR}$ and $\mathrm{AC1}$ regions along the SAMBA transect. However, 
these EWS are highly influenced by changes in properties of the noise \cite{Boers2021} which may be 
problematic in their consistency.  Unlike $\mathrm{VAR}$ and $\mathrm{AC1}$,  the restoring rate (see Methods) 
$\mathrm{RES}$ is less influenced by the properties  of the noise,  making it a more robust statistical indicator 
for critical  slowdown detection  \cite{Boers2021}.  Note that according to EWS theory (see Methods),  
$\mathrm{RES}$  decreases when approaching a saddle-node  bifurcation and reaches zero from below at the
tipping point.  Values of  $R^S_\mathrm{RES}$ and $R^T_\mathrm{RES}$  (Figures~\ref{fig:Figure_2}c,f) are 
significantly   smaller than 1 (again $p(R^X_\mathrm{RES} < 1) > 0.9$) at intermediate depths in the center 
of the SAMBA section and near  the bottom in the west.  There are a few regions along the SAMBA transect 
where all three EWS ratios are (significantly) indicating  a critical slowdown and  they are mainly found below 
500~m depths. This result is robust when slightly varying  the  latitude of the section in the South Atlantic 
(results not shown).  

There are almost no regions in the North  Atlantic  where all three EWS ratios show indications of critical 
slowdown.  EWS ratios for the RAPID transect (26$^{\circ}$N, Figure~\ref{fig:Figure_S1}) and OSNAP transect 
(60$^{\circ}$N, Figure~\ref{fig:Figure_S2}) suggest that the statistical indicators along these measurement 
transects are not effectively detecting or indicating a critical slowdown when approaching the AMOC collapse
in the CESM.  A possible explanation might be that the signal to noise ratio is  too low such that the signal is not 
captured well enough. 
Basin-wide  plots of the ratios $R^S_{\mathrm{RES}}$ (Figure~\ref{fig:Figure_S3}) and $R^T_{\mathrm{RES}}$ 
(Figure~\ref{fig:Figure_S4}) at different depths  indicate substantially more significant regions in the South Atlantic
than  in the North Atlantic. From the equilibrium simulations $E_1$ and $E_2$ it can therefore be concluded 
that values of $R^S_{\mathrm{RES}}$ and $R^T_{\mathrm{RES}}$ along the 34$^{\circ}$S transect are 
the most effective in measuring a critical slowdown of the AMOC. 

Earlier analysis \cite{Boulton2014, Feng2014} also focused on classical EWS using data from a FAMOUS 
model simulation \cite{Hawkins2011},  which showed an AMOC collapse under a relatively rapid freshwater forcing 
change with a rate of 5 $\times$ 10$^{-4}$ Sv yr$^{-1}$.  The results of \cite{Boulton2014} show that the 
variance and lag-1 autocorrelation in the annual resolution data are most reliable in the high northern latitudes 
and at the southern boundary of the Atlantic. However, the analysis of \cite{Feng2014} indicates no early 
warning signs in the variance and lag-1 autocorrelation  using the same model, which is more in agreement 
to the analysis provided here. Although \cite{Feng2014} uses a different method than \cite{Boulton2014} and 
averages the data over the latitudes,  they only find a strong anomalous signal in the kurtosis indicator when 
combining AMOC data along several latitudinal transects in the North and South Atlantic, including the RAPID 
and SAMBA transects. Our equilibrium analysis suggests a robust EWS  along one transect 
only, namely the 34$^{\circ}$S transect, making it a more easily computable  EWS compared to the 
complex network based one  in \cite{Feng2014}. 

\section*{Tipping Times}

We next test whether data of  the 34$^{\circ}$S transect of  the quasi-equilibrium CESM  simulation can be used  
to determine the tipping time, i.e. the time that the AMOC collapses.  Here, we assume,  based on the EWS theory, 
that the tipping time $\tau_e$ is associated  with a zero restoring rate  (see Methods). First, we determine the local  
restoring rate over the quasi-equilibrium  simulation, with a sliding window of 70~years where the data is linearly  
detrended. The restoring rate time  series are limited to model year~1635, i.e. the last sliding window covers  
model years~1600 -- 1670. In this way, the AMOC collapse is excluded from the analysis. Next, we determine 
the  change point  (CP), based on a change in the statistical properties of the  given time series (see Methods),  in 
the interval between model years~1300 and 1600. To address the robustness of the CP analysis,  we allow  the CP 
to vary between model year~1300 and CP$_{\mathrm{end}}$, where CP$_{\mathrm{end}}$ varies between 
model year 1500 and 1600. Data from the CP to model year~1635 are used to linearly fit the restoring  rates 
which are then extrapolated to zero to find $\tau_e$ (Figures~\ref{fig:Figure_S5} and \ref{fig:Figure_S6}). 

\begin{figure}[h!]

\includegraphics[width=1\columnwidth, trim = {0cm 0.3cm 0cm 0.3cm}, clip]{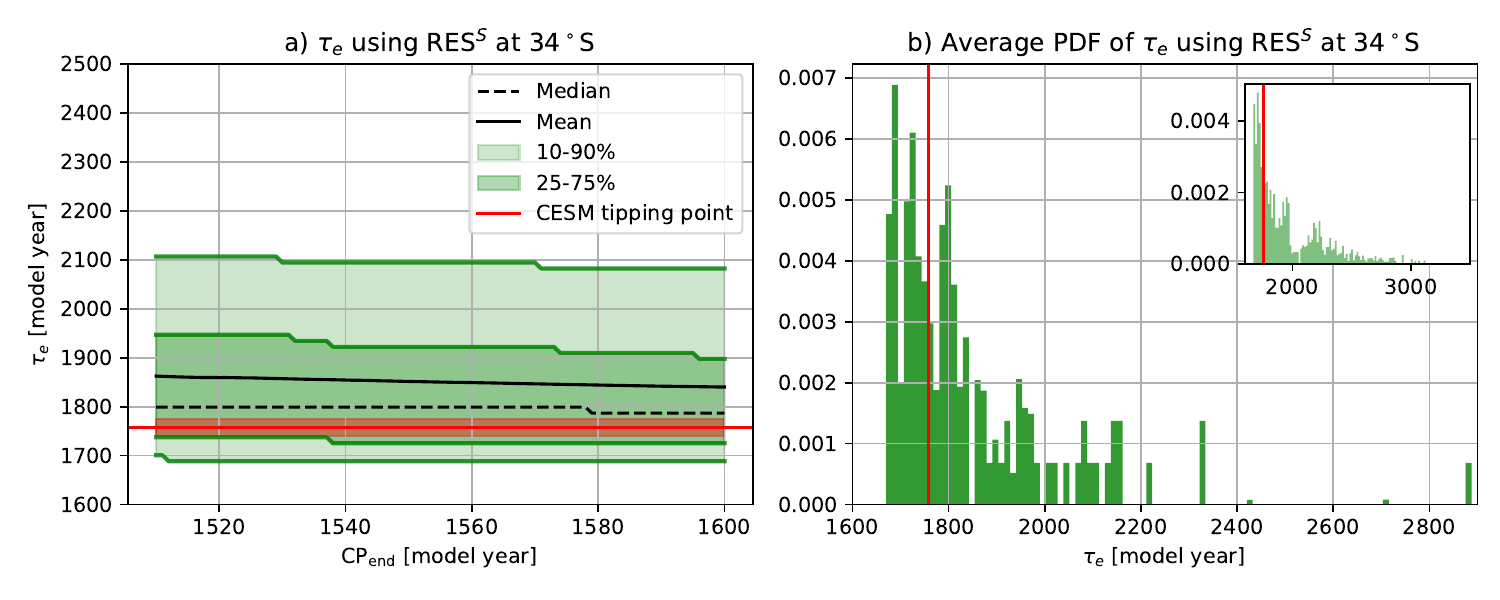} \\

\includegraphics[width=1\columnwidth, trim = {0cm 0.3cm 0cm 0.3cm}, clip]{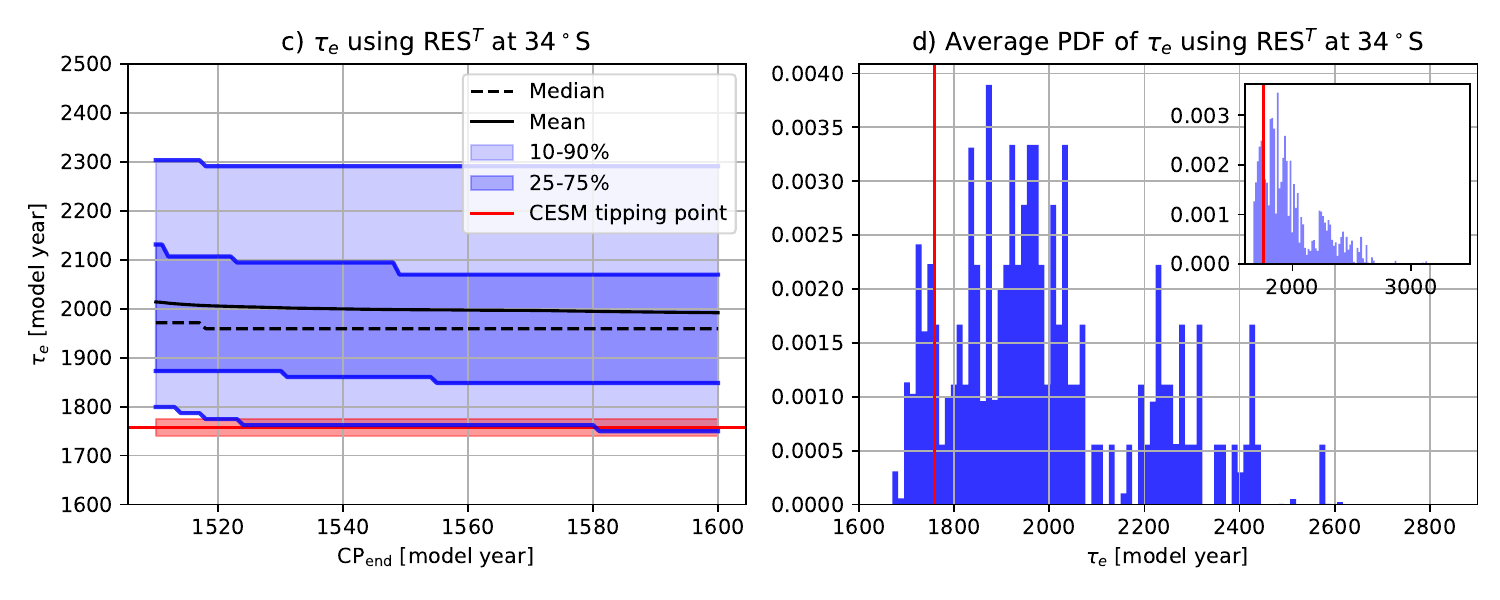}

\caption{\textbf{Estimated AMOC tipping time $\tau_e$ along 34$^{\circ}$S.}
(a): The estimated AMOC tipping time based on the salinity restoring rates of the quasi-equilibrium simulation along the SAMBA 
transect at 34$^{\circ}$S. For each local significant $R^S_{\mathrm{RES}}$ (Figure~\ref{fig:Figure_2}c), we determine the 
AMOC tipping time  for varying CP$_{\mathrm{end}}$. For each CP$_{\mathrm{end}}$, we show the PDFs, including 
the mean, median, and the confidence intervals. The actual AMOC tipping time in model year~1758 (1741 -- 1775, 
10\% and 90\% percentiles, respectively) is indicated in red. 
(b): The tipping time PDF for all CP$_{\mathrm{end}}$ (panel~a), the red line is the AMOC tipping time of model year~1758.
The inset shows a similar PDF, but consists of all (i.e., significant and non-significant $R^S_{\mathrm{RES}}$) grid points along the SAMBA transect.
(c \& d): Similar to panels~a and b, but now for the temperature restoring rates and we use the local 
$R^T_{\mathrm{RES}}$ (Figure~\ref{fig:Figure_2}f).}

\label{fig:Figure_3}
\end{figure}

A linear fit at a particular grid point is only included for estimation of the tipping time when 
the time series increases (significantly) monotonically (see Methods) and when the 
$R^S_{\mathrm{RES}}$ value (as determined from the simulations $E_1$ and $E_2$)  
is significant (Figures~\ref{fig:Figure_2}c,f). The PDFs of the tipping times using  these 
grid points are shown in Figures~\ref{fig:Figure_3}b,d. We find  that the median of the estimated 
AMOC tipping time (Figure~\ref{fig:Figure_3})   is model year~1787 (1688 -- 2082, 10 and 90\% 
percentiles, respectively) for the largest CP$_{\mathrm{end}}$ value. Using a similar  approach, 
\cite{vanWesten2024} found the actual  AMOC tipping time at  model year~1758  (1741 -- 1775, 
10\% and 90\%  percentiles, respectively).  Hence, a reasonable AMOC tipping time estimate is
found  using the locations for which $R^S_{\mathrm{RES}}$ is significant.  The actual AMOC tipping 
time falls inside the  25\% to 75\% percentile and is robust for varying CP$_{\mathrm{end}}$.  
However, for  temperature, using $R^T_{\mathrm{RES}}$ in a similar way to determine the 
grid points for the linear fits,  this   estimate is  model year~1959  (1750 -- 2291, 10 and 90\%  
percentiles, respectively) and hence is not close to the AMOC tipping time. 

When we consider all (i.e., significant and non-significant $R^S_{\mathrm{RES}}$) 
grid points for the estimate and apply the same criteria for the fits, the tipping time  PDF shifts to later 
years (inset in Figure~\ref{fig:Figure_3}b).  The actual AMOC tipping time still falls inside the 25\% to 
75\% percentile and is also robust for varying CP$_{\mathrm{end}}$ (not shown). 
The estimate from temperature restoring rates is improving slightly when we 
consider all grid points (inset in Figure~\ref{fig:Figure_3}d).   Applying the same analysis on the 
RAPID and OSNAP transects results in an inconsistent estimate for the AMOC tipping time. 
For example, most grid points (with a significant $R^S_{\mathrm{RES}}$ or $R^T_{\mathrm{RES}}$) 
show no CP in the restoring rate time series. For the grid points where a fit was obtained, the 
10\% percentile of the PDF estimate was 100~years later than the actual AMOC tipping time. 

It is interesting that a useful tipping time estimate can only be obtained using data from about 
model year 1300 and the methodology fails when earlier model data are used. This is thought 
to be connected to the fact that the AMOC-induced freshwater convergence is negative 
\cite{vanWesten2024} only after model year 1300. As this freshwater convergence (approximately 
equal to $F_{\mathrm{ov}}$ at 34$^\circ$S)  is a measure of  the salt-advection feedback, 
the AMOC starts to decrease due to internal feedbacks only after model year 1300.  
Because from observations the present-day AMOC has a negative freshwater convergence
\cite{Weijer2019,  vanWesten2024b}, we next perform a comparable analysis on reanalysis 
data.

Using the physics-based observable $F_{\mathrm{ov}}$ at 34$^{\circ}$S \cite{vanWesten2024}, it turned 
out to be impossible to estimate the tipping time from reanalysis data.  The approach above, however, provides 
a new way to estimate such a tipping time and we apply it next on salinity data along the SAMBA transect 
for the reanalysis products ORAS5, GLORYS and SODA. The ORAS5 reanalysis dataset runs from 1958 to 
2023, and to address robustness we vary CP$_{\mathrm{end}}$ between 1978 to 2017.  The restoring rate 
is determined using a sliding  window of 10~years. Slightly shorter (more noise) and longer window (data 
limitation) lengths give similar results but due to the short noisy time series cannot be varied much. We 
use all section data in the estimation of the tipping time, as otherwise a too small number of data points 
would remain. 

\begin{figure}[h!]

\includegraphics[width=1\columnwidth, trim = {0cm 0.5cm 0cm 0.3cm}, clip]{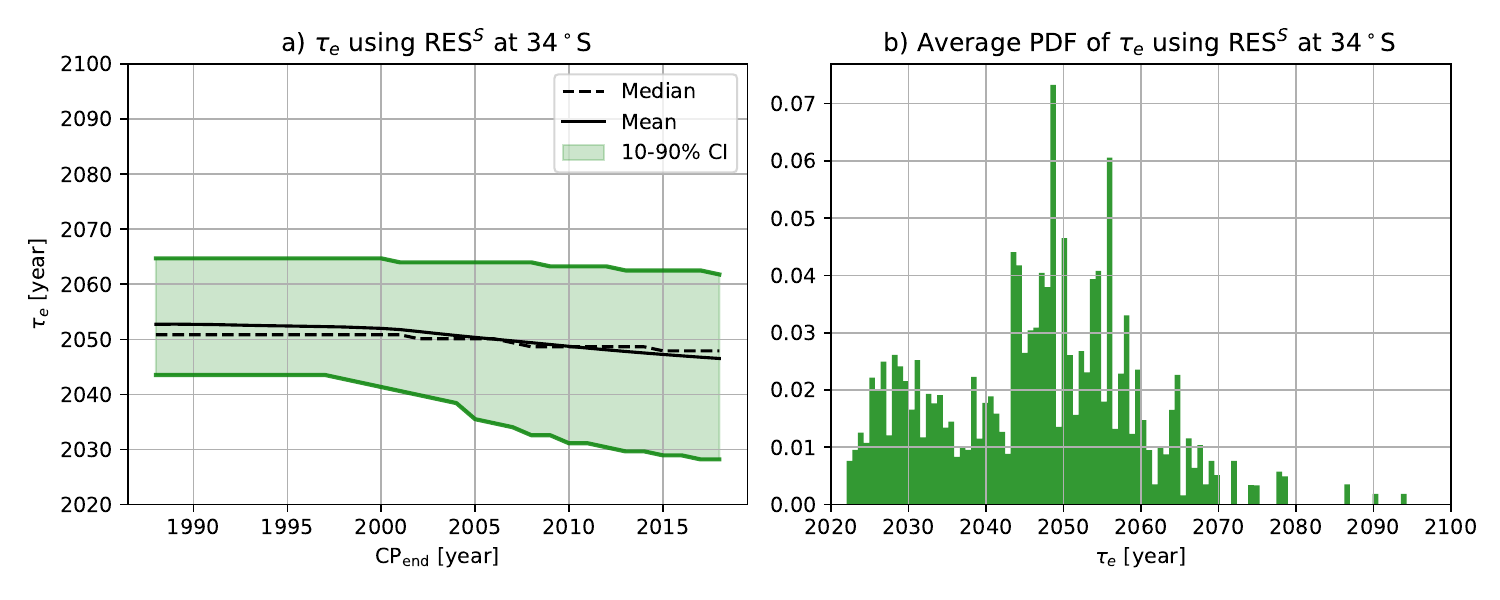}

\caption{\textbf{Estimated AMOC tipping time  along 34$^{\circ}$S in the reanalysis product ORAS5.}
Similar to Figures~\ref{fig:Figure_3}a,b, but now for the reanalysis product ORAS5 and all the grid 
cells along the SAMBA transect at 34$^{\circ}$S. } 

\label{fig:Figure_4}
\end{figure}

The mean AMOC tipping time estimate from ORAS5 is year~2050 and is robust  to  varying CP$_{\mathrm{end}}$ (Figure~\ref{fig:Figure_4}a).  The earliest year (mean 10\% percentile level) for a potential AMOC collapse is 2037 and the latest year (mean 90\% percentile level) is  2064. The average probability of an AMOC collapse before the year~2050 is $59\%$ with a standard deviation of 17\% for ORAS5 (cf. Figure~\ref{fig:Figure_4}b). Applying the same procedure to SODA and GLORYS results in 91\% and  92\% collapse  probabilities before 2050, respectively (Figure~\ref{fig:Figure_S7}). Note that these latter two  reanalysis  data sets have an even shorter length than ORAS5 and are therefore less reliable. 

\section*{Discussion}

In the IPCC-AR6 report, the probability of an AMOC collapse is considered to be low with medium confidence \cite{IPCC2021}. 
Our analysis  provides a first probability estimate from reanalysis data which gives a mean tipping time estimation of 2050 with a 
10 -- 90\% CI of  2037 -- 2064. This is comparable to the findings of \cite{Ditlevsen2023} who used the sub-polar SST index to 
estimate the AMOC tipping time to be at 2057, with a 95\% confidence interval 2025 -- 2095. Interestingly, the sub-polar SST 
index does not give an early warning signal for the AMOC collapse in the CESM quasi-equilibrium simulation \cite{vanWesten2024}
and the sub-polar gyre is also not identified here as an optimal observational region. Considering the problems with EWS 
detection using proxy based AMOC time series  \cite{Ben-Yami2023}, the tipping time estimate correspondence may just be 
coincidental. To establish the results presented in this paper, several assumptions were made, which require further 
justification. 

First, the CESM quasi-equilibrium simulation showing the AMOC collapse 
is for a pre-industrial situation with relatively high freshwater forcing.  This is obviously far outside the parameter regime of historical 
observations and mainly due to biases in this model \cite{vanWesten2024b},  in particular in the Indian Ocean freshwater 
fluxes.  We assume that for developing the optimal observation regions, the different background states do not matter. 
This is plausible as the physical mechanism of the collapse, and the associated physical variables involved, are independent 
of background conditions. This justifies our methodology of determining the tipping  time distribution from  reanalysis 
data.  

Second, the quality of the reanalysis data is questionable  for determining the collapse time distribution because these
products also depend on models that  have their biases. Furthermore, the time series are relatively short compared to the 
SST based AMOC reconstructions \cite{Ceasar2018}. On the positive side, the reanalyses show consistently 
lower biases  compared to real observations than global climate models, in particular those related to the AMOC, such as the
$F_{\mathrm{ov}}$ at 34$^{\circ}$S  \cite{vanWesten2024}. Although the assumption that reanalyses data are 
adequate here for tipping time estimation can not be fully justified, they are at the moment the best observational 
products  which are available.  

Third, the analysis provided here assumes that by extrapolating the restoring rate to zero, one can find the location 
of the tipping time of the system. While this is theoretically true when the rate of forcing is much smaller than the 
equilibration time scales of the AMOC, in both the CESM simulation and the observations there will  be an overshoot 
present depending on the forcing rate \cite[]{Ritchie2021}.  Furthermore, the  overshoot can 
be different in the reanalysis data compared to the  quasi-equilibrium simulation, giving an additional uncertainty.  
Also because we use all section grid points and not only significant ones for
the reanalysis data (cf. Figure~\ref{fig:Figure_3}c),  our   method can only provide a lower  bound of the tipping time and it is difficult to  determine 
an uncertainty measure.  Additionally,  the future forcing can be highly non-linear \cite{Riahi2017},  which can also 
affect the tipping time. 

Although there has been criticism regarding the estimation of tipping times from observational data and the 
various assumptions inherent in the estimation procedures \cite{Ben-Yami2023}, our method presented here 
offers a more physically based approach to identify optimal locations for EWS and establishing a lower bound 
of AMOC tipping times. By focussing on  the restoring rate in our critical slowdown analysis, we address the 
issues of the possibly non-stationary and/or non-white noise forcing of the AMOC \cite{Boettner2022}. 

Our analysis of the CESM results indicates that the SAMBA ($34^{\circ}$S)  transect data, in particular 
the salinity, are most useful for providing (and improving the current) estimates of  AMOC tipping probabilities. 
This result is consistent  with the recently \cite{vanWesten2024} identified physics-based indicator of an 
AMOC collapse ($F_{\mathrm{ov}}$ at 34$^{\circ}$S).  Our work therefore leads to two major conclusions. 
Observations at the southern  boundary of the Atlantic appear crucial for early warning of an AMOC 
collapse.  As a consequence, the SAMBA measurements are important to continue over the next few decades. 
Second, the probability of an AMOC collapse before the year 2100 is very likely to be underestimated in the 
IPCC-AR6 and needs to be reconsidered in the IPCC-AR7.  

\newpage
\backmatter

\section*{Methods}

\bmhead{Climate Model Simulations}

We use simulation results of the CESM version 1.0.5 (the f19$\_$g16 configuration) with horizontal resolutions of 1$^\circ$ for the  ocean/sea-ice and 2$^\circ$ for the atmosphere/land components. 
In \cite{vanWesten2023}, results from a quasi-equilibrium simulation were presented, which was branched off from the pre-industrial CESM control simulation at model year 2,800 \cite{Baatsen2020}. 
The quasi-equilibrium simulation was performed by linearly increasing the surface freshwater forcing between latitudes 20$^\circ$N and 50$^\circ$N with a rate of 3 $\times$ 10$^{-4}$ Sv yr$^{-1}$ up to model year 2,200, where it reaches a freshwater flux forcing of $F_H$ = 0.66 Sv. 
The freshwater flux anomaly was globally compensated to conserve salinity. 
From the quasi-equilibrium simulation, two new simulations were performed: one starting at model year 600 under constant $F_H$ = 0.18 Sv and one starting at model year 1500 under constant $F_H$ = 0.45 Sv. 
These simulations were continued for 500~model years where the states are in near statistical equilibrium.

\bmhead{The Freshwater Transport}

The freshwater transport of the overturning component ($F_{\mathrm{ov}}$) at latitude $y$ 
and time $t$ is determined as: 
\begin{equation}
F_\mathrm{ov}(y,t) = - \frac{1}{S_0} \int_{-H}^{0} \left[ \int_{x_W}^{x_E} v^*(x,y,z,t) \mathrm{d} x \right] \left[ \langle S \rangle(y,z,t) - S_0 \right] \mathrm{d}z
\end{equation}
where $S_0 = 35$~g~kg$^{-1}$ is a reference salinity. The $v^*$ is defined as $v^* = v - \hat{v}$,
where $v$ is the meridional velocity and $\hat{v}$  the full-depth section spatially-averaged meridional velocity.
The quantity $\langle S \rangle$ indicates the zonally-averaged salinity. \\

\bmhead{The Restoring Rate}

Classical EWS are based on a representation of the behaviour of perturbations on a statistical equilibrium state under white noise near a saddle node-bifurcation. This is defined by an Ornstein-Uhlenbeck process and can in the one-dimensional case be expressed by:
\begin{equation}
dX_t= - \lambda X_t dt + \sigma d\eta_t
\end{equation}
 where $X_t$ represents the time-dependent state variable, $\lambda$ the restoring rate, $\sigma$ the variance of the noise and $\eta_t$ the noise process. The restoring rate $\lambda$ characterises the resilience of the system, and as the system moves towards a tipping point, the restoring rate will decrease. A negative value of $\lambda$ represents a stable system state, and a saddle-node 
bifurcation point can be marked as the point where $\lambda$ reaches zero from below. 

From a discrete time series with sampling time $\Delta t$, the stationary variance (VAR)  and lag-1 autocorrelation (AC1) in case $\eta$ is a white noise process are given by $\sigma^2/(1 - \alpha^2)$ and  $\alpha$, respectively, where $\alpha = e^{-\lambda \Delta t}$.  Hence, when $\lambda \rightarrow 0$, $\alpha \rightarrow 1$ and the variance will become unbounded.  
However, in many time series, variations in variance and autocorrelation can also be related to increasing variance and autocorrelation of the external noise $\eta$ that forces the system, unrelated to critical slowdown, and therefore a modified EWS is used in \cite{Boers2021}. 
The restoring rate $\lambda$, indicated by RES, which directly quantifies the  stability of the system, can be inferred from a regression of $dX_t/dt$ onto $X_t$ under the assumption of autocorrelated residual noise with the autoregression coefficient as a free parameter. 
In this way, the estimation of RES is insensitive to increasing variance and autocorrelation of the noise and provides a robust indicator for a system approaching a saddle-node. We  therefore determine, in addition  to the variance and lag-1 autocorrelation, also the restoring rate of the time series using the procedure of \cite{Boers2021}. \\ 

\bmhead{Change Points and Restoring Rate Fits} The change point (CP) analysis is applied to the restoring rate time series.
The CP analysis detects changes in the mean and slope of the time series.  The minimum improvement in total residual error 
is set to 1 in order to limit the amount of returned change points and only get the most pronounced ones of the time series. 
The data is linearly fitted between the CP and the remaining part of the time series. A Kendall-tau test with 1000~Fourier 
surrogates is performed on the restoring rate fit and only the ones with $R > 0.7$ and $p < 0.01$ are kept.  In this way, only 
highly correlated fits which are statistically significant and increasing over time are selected. 

\bmhead{Acknowledgments}

The model simulation and the analysis of all the model output was conducted on the Dutch National 
Supercomputer Snellius within NWO-SURF project 17239. We thank Michael Kliphuis (IMAU, UU) 
for carrying out these simulations and his support in analysing the data. 

\section*{Declarations}

\begin{itemize}
\item Funding -- E.J.V.S. is funded by Utrecht University, R.M.v.W. and H.A.D. are funded by the 
European Research Council through the ERC-AdG project TAOC (project 101055096). 
\item Conflict of interest -- The authors declare no competing interest
\item Ethics approval -- Not applicable
\item Availability of data and materials -- The (processed) model output will be made available on Zenodo 
upon publication. The reanalysis and assimilation products can be accessed through: ORAS5 
(https://doi.org/10.24381/cds.67e8eeb7), GLORYS12V1 (https://doi.org/10.48670/moi-00021) and
SODA3.15.2 (http://www.soda.umd.edu).
\item Code availability -- The analysis scripts will be made available on Zenodo upon publication.
\item Authors' contributions -- E.J.V.S., R.M.v.W. and H.A.D. conceived the idea for this study.
E.J.V.S conducted the analysis and prepared all figures. All authors were actively involved in the 
interpretation of the analysis results and the writing process. 
\end{itemize}

\newpage 
\setcounter{figure}{0}

\newpage 
\setcounter{figure}{0}

\makeatletter 
\renewcommand{\thefigure}{S\@arabic\c@figure}
\makeatother


\section*{Supplementary Figures}

\begin{figure}[h]
\includegraphics[width=1\columnwidth, trim = {0cm 0cm 0cm 0cm}, clip]{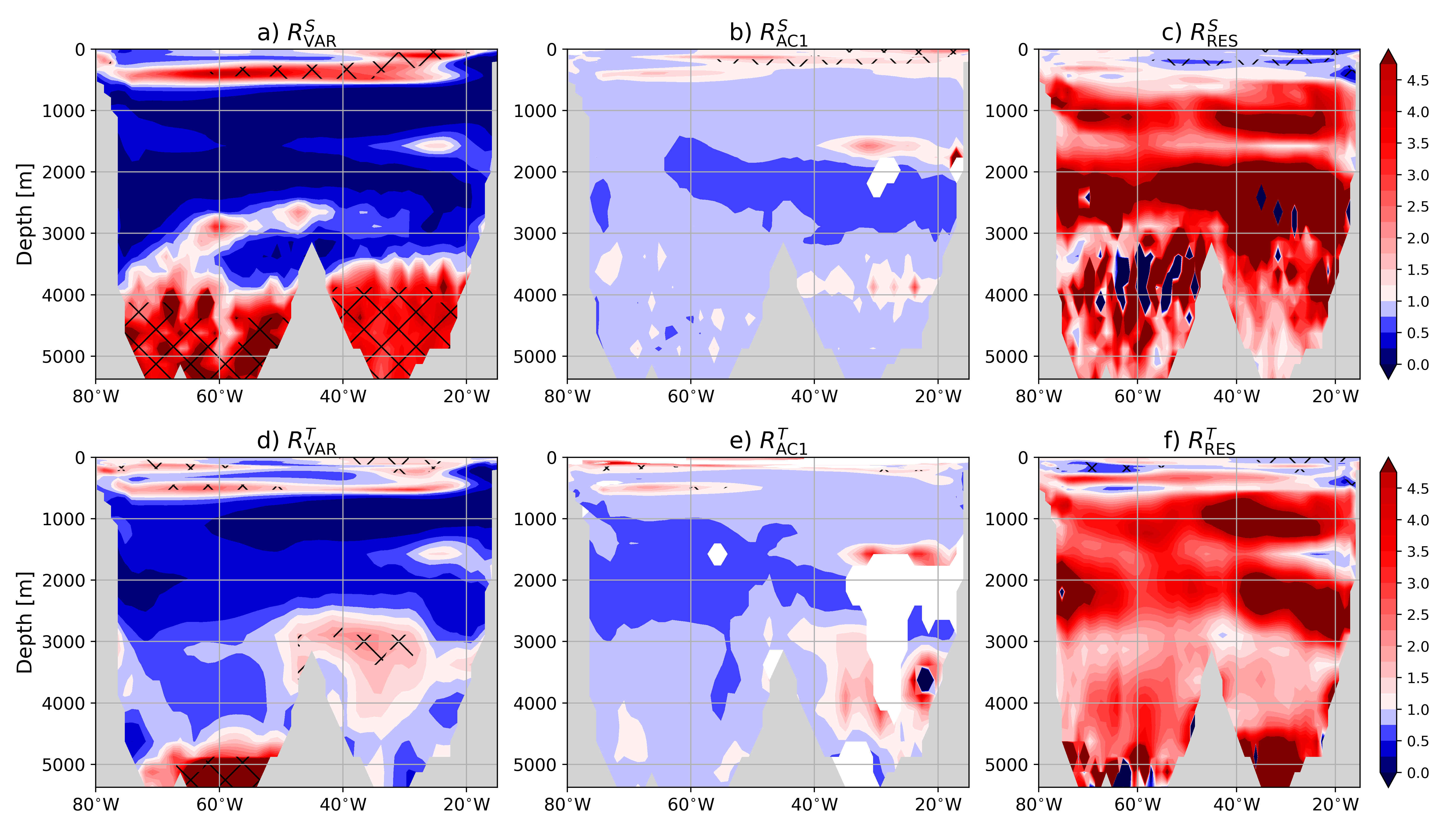}

\caption{\textbf{Early warning indicators along 26$^{\circ}$N.}
Similar to Figure~\ref{fig:Figure_2}, but now for the RAPID transect at 26$^{\circ}$N.}
\label{fig:Figure_S1}
\end{figure}


\begin{figure}[h!]
\includegraphics[width=1\columnwidth, trim = {0cm 0cm 0cm 0cm}, clip]{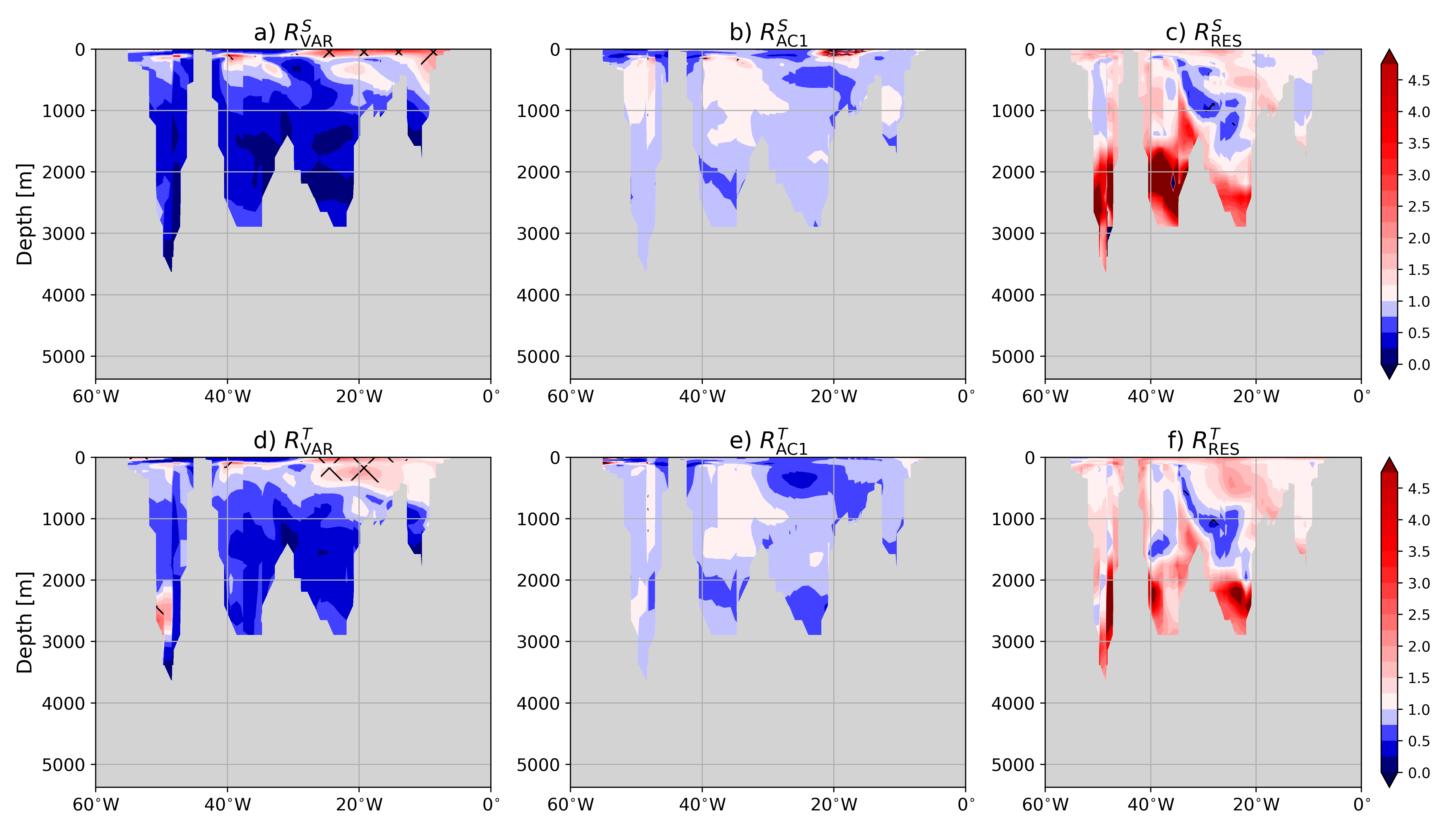}

\caption{\textbf{Early warning indicators along 60$^{\circ}$N.}
Similar to Figure~\ref{fig:Figure_2}, but now for the OSNAP transect around 60$^{\circ}$N.}
\label{fig:Figure_S2}
\end{figure}


\begin{figure}[h!]
\includegraphics[width=1\columnwidth, trim = {0cm 6cm 0cm 0cm}, clip]{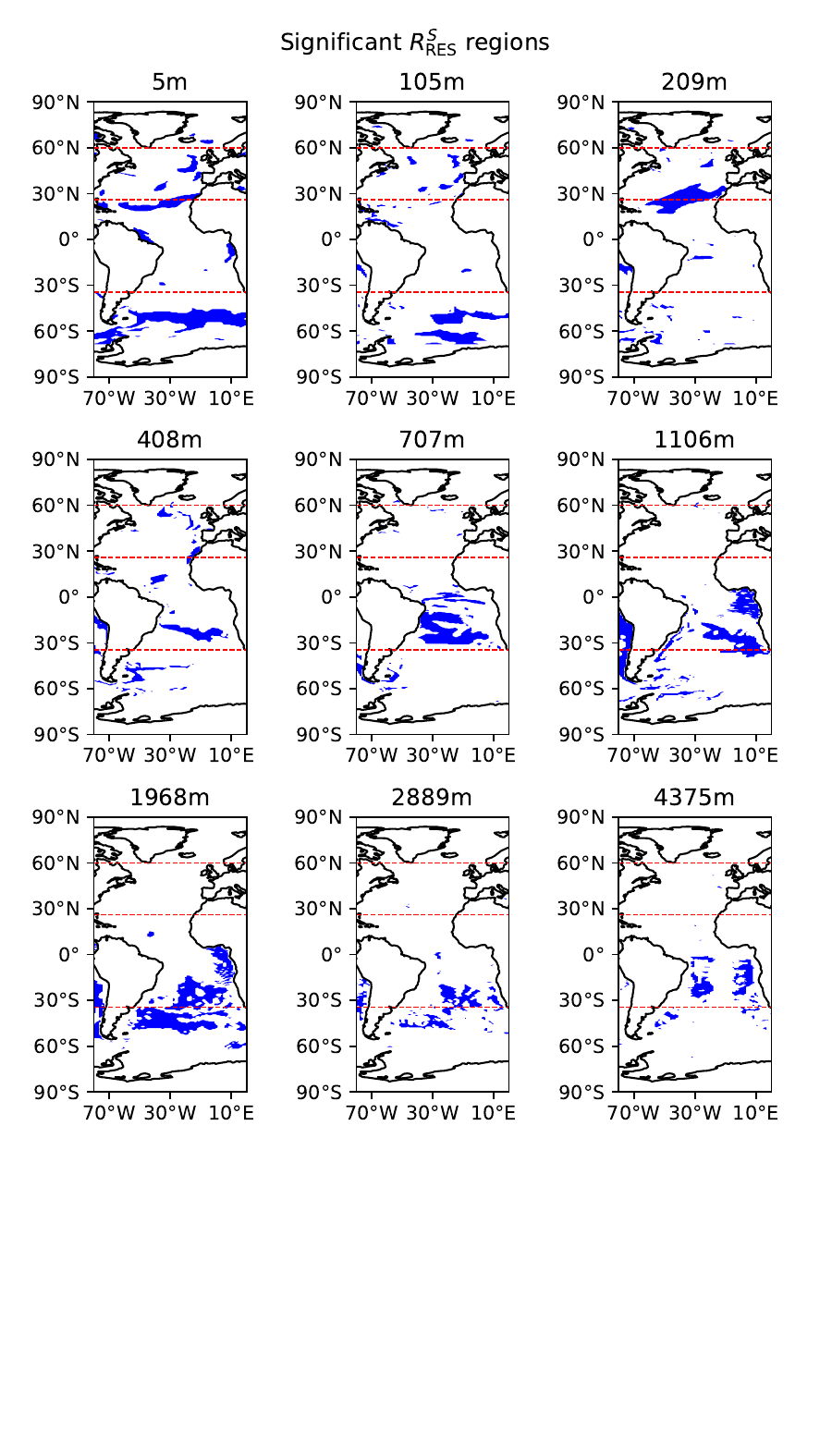}

\caption{\textbf{Early warning indicators for salinity at different depth levels.}
The blue regions indicate a significance ratio for salinity for $R^S_{\mathrm{RES}}$ and for different depth levels.
The red dashed lines indicate the transects at 34$^{\circ}$S, 26$^{\circ}$N and 60$^{\circ}$N.}
\label{fig:Figure_S3}
\end{figure}


\begin{figure}[h!]
\includegraphics[width=1\columnwidth, trim = {0cm 6cm 0cm 0cm}, clip]{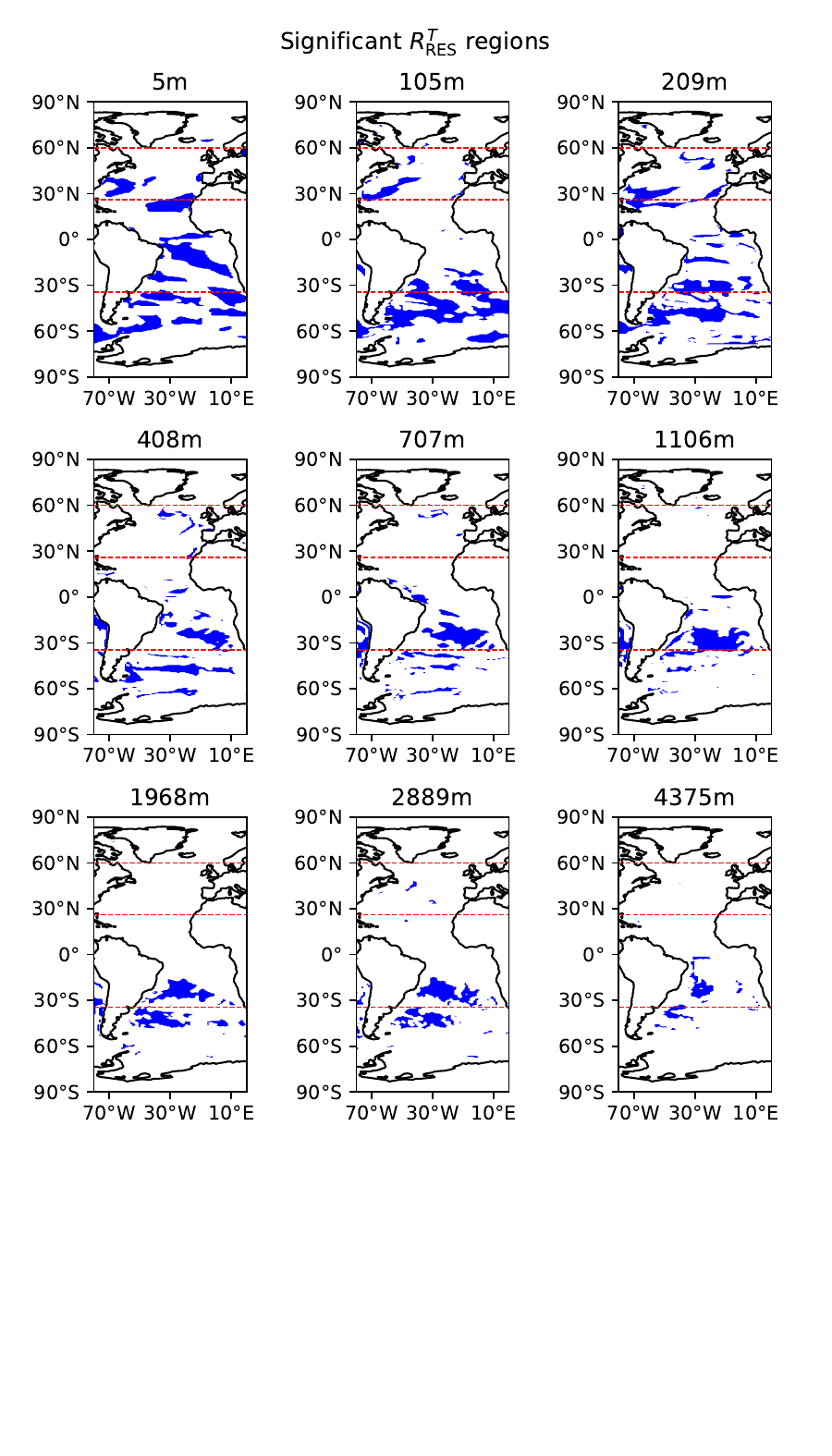}

\caption{\textbf{Early warning indicators for temperature at different depth levels.}
The blue regions indicate a significance ratio for temperature for $R^T_{\mathrm{RES}}$ and for different depth levels.
The red dashed lines indicate the transects at 34$^{\circ}$S, 26$^{\circ}$N and 60$^{\circ}$N.}
\label{fig:Figure_S4}
\end{figure}


\begin{figure}[h!]
\includegraphics[width=1\columnwidth, trim = {0cm 0cm 0cm 0cm}, clip]{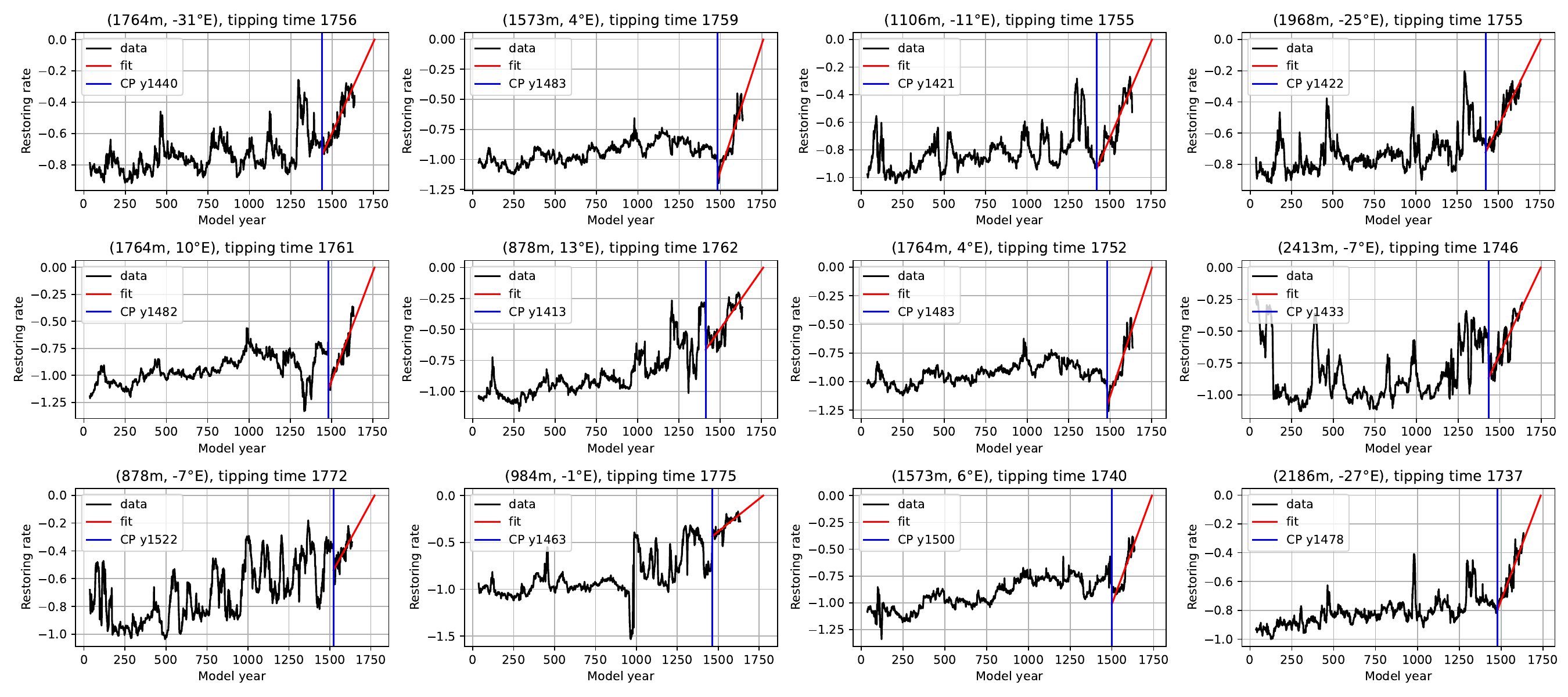}

\caption{\textbf{Estimated tipping point for the 12~best fits for salinity.}
The restoring rate for salinity (70-year sliding window) in black, the change point (CP) indicated in blue, and the 
linear fit from CP to model year~1636 and extrapolated to zero in red.
We present here the 12~best fits salinity along the SAMBA transect.
}
\label{fig:Figure_S5}
\end{figure}


\begin{figure}[h!]
\includegraphics[width=1\columnwidth, trim = {0cm 0cm 0cm 0cm}, clip]{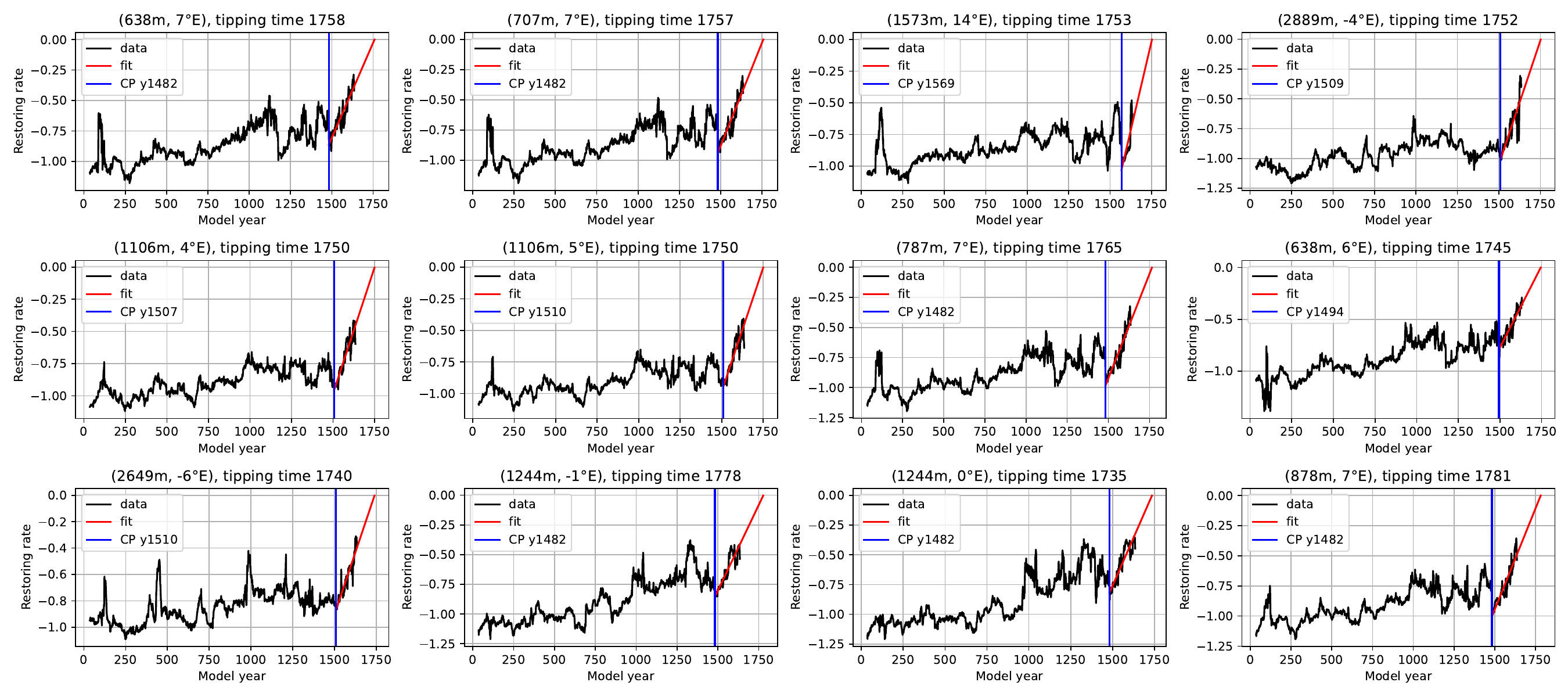}

\caption{\textbf{Estimated tipping point for the 12~best fits for temperature.}
Similar to Figure~\ref{fig:Figure_S5}, but now temperature along the SAMBA transect.}
\label{fig:Figure_S6}
\end{figure}


\begin{figure}[h!]
\begin{tabular}{c c}

\hspace{-2cm}
\includegraphics[width=0.6\columnwidth, trim = {0cm 0cm 0cm 0cm}, clip]{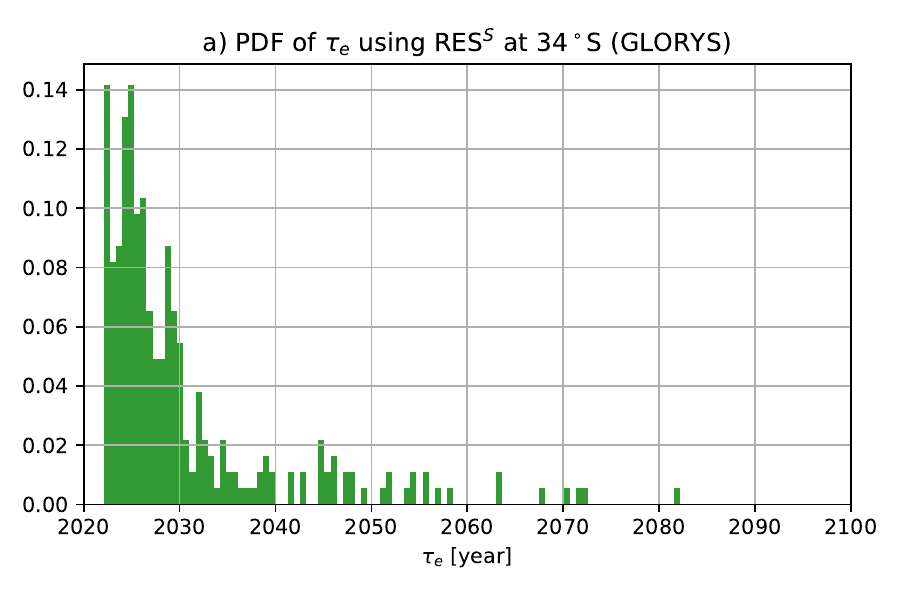} &
\includegraphics[width=0.6\columnwidth, trim = {0cm 0cm 0cm 0cm}, clip]{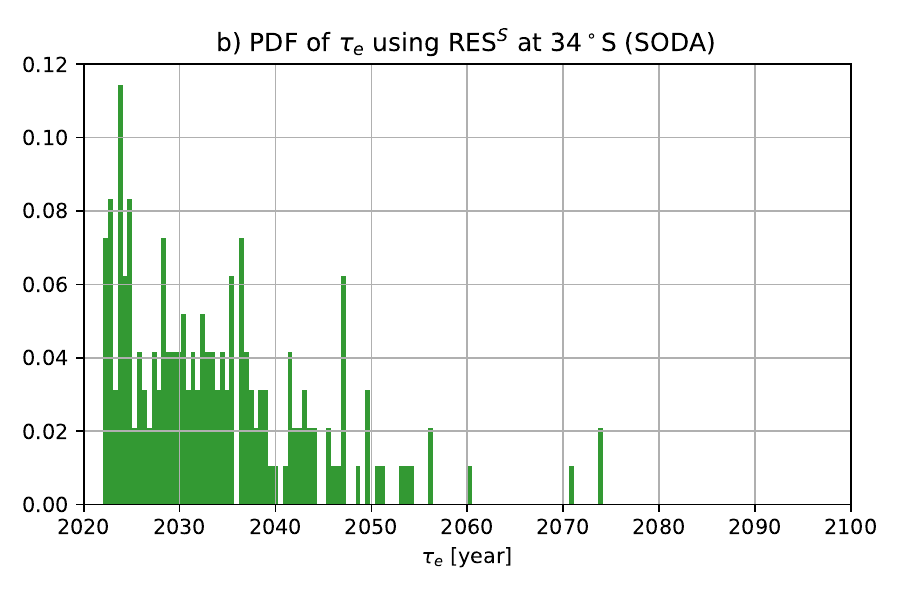}

\end{tabular}

\caption{\textbf{Estimated AMOC tipping point along 34$^{\circ}$S in reanalysis products GLORYS and SODA.}
Similar to Figure~\ref{fig:Figure_4}b, but now for reanalysis products (a): GLORYS (1993-2020) and (b): SODA (1980-2020).
}
\label{fig:Figure_S7}
\end{figure}


\end{document}